\documentclass[aps,pra,twocolumn,showpacs]{revtex4}
\usepackage[dvips]{graphicx}
\usepackage{amsmath}

\newcommand{\Er}{\mbox{$E_{\rm r}$}}
\newcommand{\vrec}{\mbox{$v_{\rm r}$}}
\newcommand{\vg}{\mbox{$v_{\rm g}$}}
\newcommand{\hk}{\mbox{$\hbar k$}}
\newcommand{\um}{\mbox{$\mu\mathrm{m}$}}

\newcommand{\us}{\mbox{$\mu\mathrm{s}$}}

\begin{document}

\title{Transport of atoms in a quantum conveyor belt}

\author{A.~Browaeys}
\altaffiliation{Institut d'Optique, Universit\'e Paris 11,  91403
Orsay, France.}
\author{H.~H\"affner}
\altaffiliation{Institut f\"ur Quantenoptik und
Quanteninformation, \"Osterreichische Akademie der Wissenschaften,
Technikerstra{\ss}e 25, A-6020 Innsbruck,
Austria.}
\author{C.~M$\mathrm{^c}$Kenzie} \altaffiliation{
University of Otago, P.O. Box 56, Dunedin, New Zealand }
\author{S.~L.~Rolston}
\altaffiliation{University of Maryland, College Park, MD 20742,
USA}
\author{K.~Helmerson}
\author{W.~D.~Phillips}
\affiliation{National Institute of Standards and Technology,
Gaithersburg, MD 20899, USA}
\date{\today}

\begin{abstract}
We have performed experiments using a 3D-Bose-Einstein condensate
of sodium atoms in a 1D optical lattice to explore some unusual
properties of band-structure. In particular, we investigate the
loading of a condensate into a moving lattice and find
non-intuitive behavior. We also revisit the behavior of atoms,
prepared in a single quasimomentum state, in an accelerating
lattice. We generalize this study to a cloud whose atoms have a
large quasimomentum spread, and show that the cloud behaves
differently from atoms in a single Bloch state. Finally, we
compare our findings with recent experiments performed with
fermions in an optical lattice.
\end{abstract}


\maketitle

An optical lattice is a practically perfect periodic potential for
atoms, produced by the interference of two or more laser beams. An
atomic-gas Bose-Einstein condensate (BEC)\cite{Anderson,dafolvo99}
is a coherent source of matter waves, a collection of atoms, all
in the same state, with an extremely narrow momentum spread.
Putting such atoms into such a potential provides an opportunity
for exploring a quantum system with many similarities to electrons
in a solid state crystal but with unprecedented control over both
the lattice and the particles. In particular we can easily control
the velocity and acceleration of the lattice as well as its
strength, making it a variable ``quantum conveyor belt". This
allows us to explore situations that are difficult or impossible
to achieve in solid state systems. The results are often
remarkable and counter-intuitive. For example atoms that are being
carried along by a moving optical lattice are left stationary when
the still-moving lattice is turned off, in apparent violation of
the law of inertia.

A few experiments have studied quantum degenerate atoms in moving
optical lattices~\cite{Fallani03,Morsch01,Eiermann03,Denschlag02}.
Bragg diffraction of a Bose condensate is a special case of
quantum degenerate atoms in a moving lattice~\cite{Kozuma99}.
Here, using a Bose-Einstein condensate and a moving lattice, we
achieve full control over the system, in particular its initial
quasimomentum and band index as well as its subsequent evolution.
We also show the difference in behavior when the atom sample has a
large spread of quasimomenta, as compared with the narrow
quasimomentum distribution of a coherent BEC.

Our lattice is one-dimensional along the $x$ axis, produced by the
interference of two counter-propagating laser beams, each of
wave-vector $k = 2\pi/\lambda$ ($\lambda \approx 589$ nm is the
wavelength of the laser beams). This results in a sinusoidal
potential, $V\sin^2k x$, with a spatial period $\lambda / 2$.

\begin{figure}
\includegraphics[width=8cm]{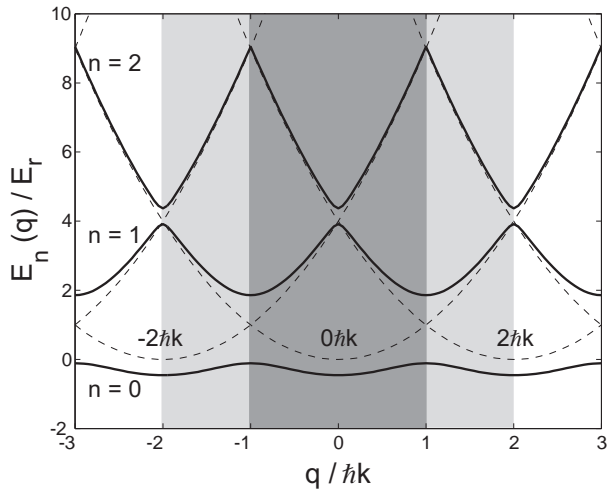}
\caption{Band structure for a 4\Er\ deep lattice in the
repeated-zone scheme. The dotted lines represent the free particle
parabolae to which the bands adiabatically connect as
$V\rightarrow 0$. The region in dark  grey corresponds to the
first Brillouin zone. The region in light grey corresponds to the
second Brillouin zone. } \label{bandstructure4Er}
\end{figure}

We will use Bloch theory, emphasizing the single particle
character of the problem. An overview of Bloch theory, as it
applies to this one dimensional system, is supplied in
reference~\cite{Denschlag02}. Briefly, the wave function of the
atoms in the lattice can be decomposed into the Bloch eigenstates
$u_{n,q}(x) e^{iqx}$ characterized by a band index $n$ and a
quasimomentum $q$, defined in the rest frame of the lattice. The
eigenenergies of the system, $E_n(q)$, as well as the eigenstates
are periodic in $q$ with a periodicity $2\hk$, the reciprocal
lattice vector of the lattice. A wave packet in band $n$ with
quasimomentum distribution centered at $q$, has a group velocity
$\vg = dE_n(q)/dq$. Fig.~\ref{bandstructure4Er} shows the band
structure in the repeated-zone scheme~\cite{Ashcroft}, for a
lattice with a depth $V =4\Er$ (\Er\ is the single-photon recoil
energy given by $\Er=\hbar^2 k^2/2M$ and is related to the recoil
velocity \vrec\ by $ \vrec =  M\vrec^2/2$, $M$ being the mass of
an atom). Note that for convenience the band-energies $E_n$ are
offset such that they coincide at large band index with the free
parabolae; this shows more clearly the avoided crossings between
free particle states due to the laser-induced coupling. These
avoided crossings create the band gaps that separate energy bands
with different indices $n$.

\section{Experimental setup}

The experimental setup has been described
previously~\cite{Kozuma99}. An almost pure Bose-Einstein
condensate   (no discernable thermal component) of about $2\times
10^{6}$ sodium atoms is prepared in a triaxial Time Orbiting
Potential (TOP) trap~\cite{Petrich95,Kozuma99}. We adiabatically
expand the condensate by lowering the mean trapping frequency
\footnote{The magnetic trap has trapping frequencies $ \omega_x =
\sqrt{2}\omega_y = 2 \omega_z$. The mean oscillation frequency is
$\omega_y$. The lattice is in the  $x$ direction, gravity being
along the $z$ direction. } from 200 Hz to a value ranging from 100
Hz to 19 Hz. This reduces the atom-atom interaction, the strength
of which is given by the chemical potential $\mu = \frac{4\pi
\hbar^2 n a}{M}$~\cite{dafolvo99}, $n$ being the density at the
center of the cloud and $a\approx 2.8$ nm the scattering length.
During the expansion, the calculated Thomas-Fermi diameter, $2R$,
of the condensate along the lattice direction increases from 18
\um\ up to values ranging from  24 \um\ to 48 \um. The
wave-function of each atom thus covers more than a 100 lattice
sites and is an excellent approximation of a Bloch state. The rms
width of the momentum distribution of the atoms in the condensate
along the axis of the lattice is $\sqrt{3}\,
\hbar/R$~\cite{Stenger99}. Therefore the rms width of the
quasimomentum distribution of each atom is $\Delta q \sim
(\lambda/ 4 R)\ \hk \le 0.01 \hk$.

To form the lattice we use two counter-propagating laser beams
perpendicular to gravity. Each has a power of up to 10~mW and is
detuned either 200-350~GHz to the blue of the sodium D2 transition
(experiments of sections~\ref{dragging_moving_lattice}
and~\ref{single_bloch_acceleration}) or 700~GHz to the red of the
D1 transition (last section). They are focused to a beam waist of
about 200~$\mu$m FWHM, leading to a calculated spontaneous
emission rate $\le 30 \ \rm{s}^{-1}$, negligible during the time
of the experiments. The lattice depth, measured by observing the
Bragg diffraction~\cite{Denschlag02}, is up to 13\Er. We use
acousto-optic modulators to independently control the frequencies
and intensities of the beams. The unmodulated intensity is kept
constant to within 5\% by active stabilization. A frequency
difference $\delta$ between the two beams produces a ``moving
standing wave" of velocity $v = \delta/2k$. Numerically, a
difference of $\delta/ 2 \pi = 100$ kHz corresponds to a lattice
velocity of one recoil velocity,  $\vrec =\hk/M \approx 3$ cm/s.

The cloud's momentum  is analyzed using time-of-flight. The
time-of-flight period, typically a few milliseconds, converts the
initial momentum distribution into a position distribution, which
we determine using near-resonance absorption imaging along an axis
perpendicular to the axis of the lattice.

\section{Dragging a condensate in a moving lattice}\label{dragging_moving_lattice}

In a first set of experiments, we begin with a BEC in a magnetic
trap with a 19 Hz mean frequency. This weak trap makes the
interactions between atoms almost negligible on the time scale of
the experiment, i.e. $\hbar /\mu$ is generally longer than the
duration of the experiment~\footnote{For this weak trap, $\mu
\approx 400$ Hz.}. After turning off the magnetic trap, we
adiabatically apply a moving lattice with a final depth of 4\Er.
The turn-on time of the lattice intensity is 200~$\mu$s, an
interval chosen to ensure adiabaticity with respect to band
excitation~\footnote{The lattice turn-on is approximately linear
in the sense that the voltage sent to the acousto-optic modulators
was linear. Their non-linear response leads to a smoother
variation of the intensity of the light, especially at the
beginning and the end of the ramp.} (see section
\ref{momentum_analysis}). The fixed velocity of the lattice, $v$,
is between 0 and about 3$\vrec$. In the lattice frame the atoms
have a quasimomentum $q = - M v$. Because the width of the
quasimomentum distribution is very narrow, this procedure produces
a good approximation of a single Bloch state with a freely chosen
$q$.

Atoms loaded in this way are dragged along with the moving
lattice. In the limit that the lattice is very deep so that the
bands are flat (i.e. $\frac{d E(q)}{dq} = 0$), the group velocity
with respect to the lattice, $\vg$, is 0 and the atoms are dragged
in the lab frame at the velocity of the lattice. For finite depth
lattices the dragging velocity in the lab frame is $v+\vg$. (Note
that for $v>0$, $\vg<0$ so that this dragging velocity in the lab
frame $v+\vg\le v$.)

\begin{figure}
\includegraphics[width=8cm]{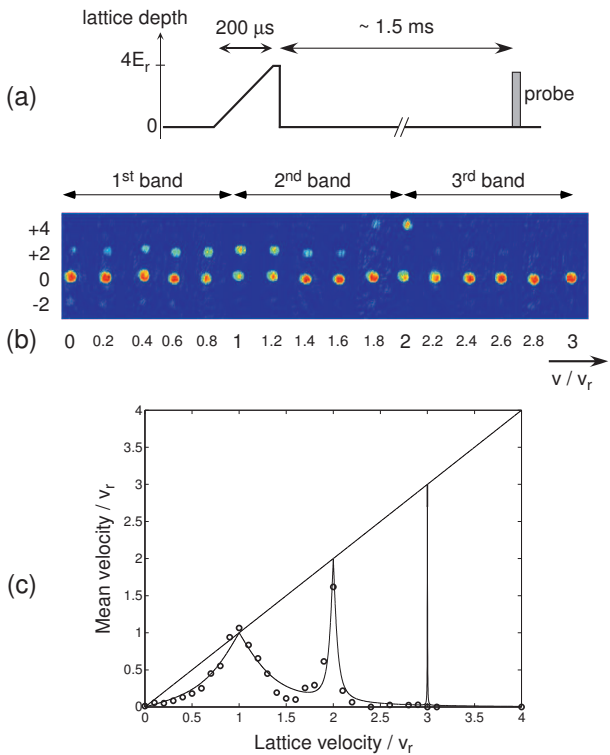}
\caption{(Color) Dragging of atoms in a moving lattice followed by
a sudden turn-off of the lattice.
Fig.~\ref{movinglatticesuddenoff}a presents the time sequence.
Fig. ~\ref{movinglatticesuddenoff}b shows the absorption image of
the cloud after a 1.5 ms time of flight following a sudden turn
off of the lattice for different lattice velocities $v$, related
to the quasimomentum by $q = -M v$. The numbers on the vertical
axis refer to the atomic velocity in units of \vrec. The average
velocity of the atoms in the lab frame, deduced from
fig.~\ref{movinglatticesuddenoff}b, is shown in
fig.~\ref{movinglatticesuddenoff}c versus the velocity of the
lattice. The initial velocity of the condensate in the magnetic
trap fluctuates with an RMS value of 0.03\vrec. The mean velocity,
after suddenly turning off the lattice, thus exhibits the same
fluctuations. The solid curve is the mean velocity of the atoms
calculated from the band structure for a 4\Er\ deep lattice. }
\label{movinglatticesuddenoff}
\end{figure}

In order to experimentally measure the dragging velocity we
suddenly (on the order of $200$ ns) turn off the moving lattice,
projecting the Bloch state onto the basis of free-particle
momentum eigenstates while preserving the momentum distribution.
Figure~\ref{movinglatticesuddenoff}a shows the lattice depth as a
function of time. Images of the resulting diffraction pattern for
various lattice velocities are presented in
figure~\ref{movinglatticesuddenoff}b.
The average velocity seen from the diffraction pattern (the
weighted average of the velocities of the individual diffraction
components) increases with the lattice velocity through the first
Brillouin zone. In fact for this rather flat band the dragging
velocity is roughly equal to the lattice velocity. (The details
for higher velocities are discussed in the following section.)

\begin{figure}
\includegraphics[width=8cm]{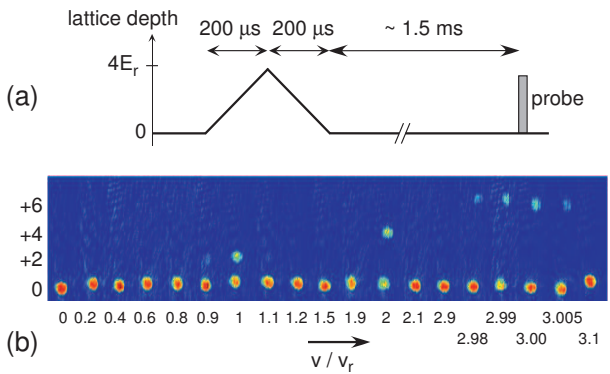}
\caption{(Color) Dragging of atoms in a moving lattice followed by
an adiabatic turn-off of the lattice.
Fig.~\ref{movinglatticeadiaboff}a presents the time sequence.
Fig.~\ref{movinglatticeadiaboff}b shows the absorption image of
the cloud after a 1.5 ms time of flight following the adiabatic
turn-off of the lattice for different lattice velocities. The
numbers on the vertical axis refer to the atomic velocity in units
of \vrec.} \label{movinglatticeadiaboff}
\end{figure}

An alternate method to study the atomic momentum is to release the
condensate adiabatically ($\sim 200\ \mu$s) rather than suddenly,
thus avoiding diffraction. Figure~\ref{movinglatticeadiaboff}a
shows the lattice intensity time sequence for this method. The
corresponding images for various lattice velocities appear in
figure~\ref{movinglatticeadiaboff}b.
These pictures show that (apart from when the lattice velocity is
very close to an integer multiple of \vrec, a situation discussed
in section~\ref{momentum_analysis}) the atoms are back at rest in
the laboratory frame, despite the fact that the lattice is still
moving during the ramping down of its intensity. This is true even
in the first Brillouin zone where the lattice drags the condensate
at roughly the lattice velocity. This result is especially
surprising when one considers that atoms moving with the lattice
return to zero velocity as if they had no inertia. One might also
ask how do the dragged atoms ``know" that they should be at rest
when the lattice is turned off. One way of understanding this is
to note that the lattice turns on adiabatically and turns off
adiabatically along the same path. This must necessarily return an
eigenstate of the hamiltonian to the same eigenstate. A more
detailed explanation involving band-structure will be presented in
the next section.

\section{Analysis of the experiments}\label{momentum_analysis}

All experiments described in this paper start with an adiabatic
turn-on of a lattice moving at a velocity  $v$. In the lattice
frame, in the limit of a vanishingly small lattice depth, the
atomic wavefunction of a momentum eigenstate has a phase gradient
$-M v/\hbar$ corresponding to the velocity $-v$ of the atoms with
respect to the lattice. This free particle state is also a Bloch
state with a quasimomentum $q$ corresponding to a phase gradient
$q/\hbar$, so that $q = -M v$.  All changes in the lattice
intensity preserve this quasimomentum (as can be seen by
calculating that the  matrix elements of the periodic potential
between Bloch states of different $q$, are zero). When the lattice
is fully turned on, the quasimomentum is still $-M v$ and if the
turn-on has been adiabatic (so that no change in band index
occurs), we end up in a single Bloch state.

Referring to figure~\ref{bandstructure4Er} we see that, when atoms
are loaded adiabatically into the lattice with the quasimomentum
in the first Brillouin zone, the free particle momentum connects
to the corresponding quasimomentum in the lowest, $n=0$, band. For
quasimomenta outside the first Brillouin zone, the free particle
momenta connect to the corresponding quasimomenta in the
appropriate band. For example if the velocity of the lattice is
$1.5 \vrec$, i.e. in the second Brillouin zone, the atoms will end
up in the second, $n=1$, band with a quasimomentum $q = -1.5 \hbar
k$. There is thus a strict relation between the range of
quasimomenta and the band index into which the atoms are loaded:
if the quasimomentum is in the $n^{\rm th}$ Brillouin zone, the
atoms are loaded into band  $n-1$. On the other hand, if for
example we wish to prepare the atoms in $q = -1.5 \hbar k$ and
$n=0$, we would have to accelerate the lattice, as described in
section~\ref{single_bloch_acceleration}.

The condition for adiabaticity with respect to band excitation
during the loading has been detailed in ref.\cite{Denschlag02}: in
order to avoid transitions from a given band to an adjacent band,
the rate of change  of the lattice depth $V$ must fulfill
$\frac{dV}{dt} |\langle n, q|\sin^2 kx|n\pm 1,q\rangle| \ll \Delta
E^2/\hbar$. $\Delta E$ is the energy difference between the given
band and its nearest neighbor. When $\Delta E $ approaches 0 (as
is usually  the case near a Brillouin zone boundary when $V
\rightarrow 0$) the process cannot be adiabatic. For $q= 0, n=0$,
$\Delta E \ge 4\Er$, and the natural time scale for adiabaticity
with respect to band excitation is on the order of $h/4\Er$. We
emphasize that in the limit of $V \rightarrow 0$ there is a
natural energy gap due to the periodicity of the lattice, $\Delta
E \neq 0$ (except at the edge of the Brillouin zones). The
existence of this non-zero energy gap when the lattice depth goes
to zero is in contrast to, for example, a harmonic oscillator for
which the spacing between energy levels does go to zero as the
strength of the potential vanishes.

We now analyze in more detail the two methods for studying the
momentum distribution described in the previous section.

In the first method we turn off the lattice potential suddenly,
i.e. diabatically. This sudden turn-off leaves the atomic momentum
distribution unchanged from what it was in the lattice. If the
atoms are in a Bloch state, corresponding to a single value of
$q$, the wave-function as viewed in the rest frame of the lattice
is a superposition of plane waves with momenta $q + 2 m \hbar k$
($m$ is an integer). The population-weighted average of the
momentum components gives the mean momentum of the atoms in the
lattice, which is $M \vg$~\cite{Ashcroft}. In the laboratory frame
these momentum components are shifted by the velocity of the
lattice and are observed as a diffraction pattern. The
time-of-flight spatial distribution of these momentum components
is analogous to the diffraction pattern of any wave from a
periodic structure. The spacing between the momentum components
gives the reciprocal lattice vector, 2\hk\ in our case. This
diffraction is characteristic of sudden turn-off (or on) of the
lattice.

Figure~\ref{movinglatticesuddenoff}c shows the measured dragging
velocity in the lab frame as a function of the lattice velocity $v
= - q/M$. Also shown is the calculated dragging velocity $\frac{d
E(q)}{dq} + v$ for a 4\Er\ deep lattice. When the atoms are in the
first Brillouin zone and in the $n = 0$ band they are dragged
along at close to  the lattice velocity, because the $n = 0$ band
is nearly flat (see figure~\ref{bandstructure4Er}). The next,
$n=1$, band is much less flat and the atoms are not dragged at the
lattice velocity except at the edge of the Brillouin zone where
$\vg = \frac{d E(q)}{dq}$ vanishes. In the third Brillouin zone,
the $n=2$ band is so close to a free particle that there is almost
no dragging and experimentally we do not even see good dragging
near the zone boundary at 3\vrec\ because the feature is too
narrow. This behavior is rather intuitive in that the lattice
drags atoms effectively up to a velocity for which the atomic
kinetic energy in the lattice frame is about equal to the lattice
depth. Reference~\cite{Fallani03} reported similar results,
measuring the dragging velocity using the displacement of the
cloud rather than diffraction. (Note that they plot the group
velocity.) This dragging process is also discussed
in~\cite{Eiermann03}.

Now let us consider the rather counter-intuitive results obtained
by adiabatically ramping off the lattice intensity. As noted
earlier, turning off the lattice either adiabatically or
non-adiabatically does not change the quasimomentum distribution,
although it may change the momentum distribution. (This assumes
that no other forces besides the lattice act on the atoms in the
rest frame of the lattice. This assumption would be violated, for
example, in the presence of interaction between the atoms or if
the lattice were accelerated.)

Consider a single Bloch state in the lattice, as is the case in
the previous section. In contrast to the sudden turn-off method
described above, the multiple momentum states $q+2m\hk$ coalesce
into a single momentum component, whatever the depth of the
lattice was. Looking at figure~\ref{bandstructure4Er}, we can see
that any {\it single} Bloch state $|n,q\rangle$ will adiabatically
connect to a {\it single} free particle parabola, unless there is
a degeneracy and adiabaticity fails. For the specific experiment
described in section~\ref{dragging_moving_lattice}, where a
lattice moving at a constant velocity is turned on and off,  this
parabola is always the one labelled $0\hk$.  In this case the
Bloch state produced is such that the single momentum component is
$p = q = -Mv$ in the frame of the lattice. Transforming into the
lab frame we find the velocity of the atoms to be zero, as
observed.

As an alternate explanation we recall posing the question ``how do
the dragged atoms `know' that they should be at rest when the
lattice is turned off?". We now can see that this information is
stored in the phase gradient, or the quasimomentum, which does not
change as the lattice is ramped on and off. We again emphasize
that, in the absence of interactions, this phase information is
preserved no matter how deep the lattice was or how fast the
lattice was turned on and off.

Let us now return to the failure of adiabaticity near the edge of
the Brillouin zones. Referring to figure~\ref{bandstructure4Er},
consider free atoms, stationary in the lab frame, but  at the edge
of a Brillouin zone in the lattice rest frame, for example at $q =
\hk$ or $q = 2\hk$. At $q = \hk$ atoms will, as the lattice is
turned on, be loaded into both bands $n = 0$ and $n = 1$; at $q =
2\hk$ atoms will be loaded into $n=1$ and $n=2$. Upon turning off
the lattice, the two populated states will each connect to two
free-particle parabolae. For example at $q = \hk$ atoms will be in
both the $0\hk$ and $2\hk$ parabolae (at $q = 2\hk$, they will be
in both $0\hk$ and $4\hk$). In the lattice frame (with the lattice
off) atoms at $q = \hk$ in the $0\hk$ parabola are moving with a
group velocity $+\vrec$; atoms at $q = \hk$ in the 2\hk\ parabola
are moving with a group velocity $-\vrec$. Transforming back into
the lab frame, these atoms are moving at $0\vrec$ and $-2\vrec$
respectively. Similarly at $q= 2\hk$ in the lattice frame the
atoms are moving at $+2\vrec$ and $-2\vrec$, corresponding to 0
and 4\vrec\ in the lab frame. This is exactly what is
experimentally seen in figure~\ref{movinglatticeadiaboff}b. (And
it is exactly the same as first- and second-order Bragg
diffraction~\cite{Kozuma99}). The fraction of population in each
momentum component depends on the details of the loading and the
unloading. For higher bands the adiabaticity condition becomes
easier to satisfy near a band edge. Even though the band gap at
the level anti-crossing at the  Brillouin zone edge gets {\it
smaller} for larger band index, the energy difference $\Delta E$
between adjacent bands at a fixed distance in quasimomentum from
the Brillouin zone edge, is {\it larger} for higher bands (see
fig.~\ref{bandstructure4Er}). This larger $\Delta E$ leads to
greater adiabaticity for a given rate of change of the lattice
depth at given distance in quasimomentum from the zone edge. This
partially explains why so little population in non-zero momentum
states is seen near the band edges for high velocities in
fig.~\ref{movinglatticeadiaboff}b. In addition the coupling
between adjacent bands gets smaller for higher bands (because it
represents a higher order process), as reflected by the narrowing
of the band gap, and this smaller coupling further reduces the
population of non-zero momentum states.

This method to analyze the quasimomentum distribution by
adiabatically ramping down the amplitude of the lattice is
independent of the way this distribution has been created, and
thus allows the analysis of complex quasimomentum distributions.
In order to understand this point, we recall that there is a
unique correspondence (except at the Brillouin zone boundary)
between any given Bloch state $|n,q\rangle$ in the lattice and  a
momentum state in the lab frame when the lattice is adiabatically
turned off. For example two Bloch states with the same $q$ ($0<
q<\hk$) in a lattice moving with a velocity $v$, but in two
adjacent bands, let's say $n=0$ and $n=1$, will connect to
momentum states $q + Mv$ and $q+Mv-2\hk$ respectively. In the same
way two Bloch states with the same band index and two different
$q$'s will end up in two different momentum states. Suppose now we
prepare a given quasimomentum distribution in the lattice frame,
consisting of many $q$'s in many bands, and suppose we
adiabatically ramp down the intensity of the lattice. If during
that ramping-down time the quasimomentum distribution does not
significantly evolve (e.g. under the influence of interaction, or
under acceleration), the adiabaticity ensures that the population
in a given state of quasimomentum $q$ in band $n$ is conserved
during the process. The quasimomentum distribution is thus mapped
onto a momentum distribution in the laboratory
frame~\cite{Kastberg95}. This method, which has also been used in
reference~\cite{Greiner2001}, then allows us to fully reconstruct
the quasimomentum and band distribution.

We will give other examples of such  mappings in the next two
sections.

\section{Acceleration of a condensate in a single Bloch
state}\label{single_bloch_acceleration}

In this section, we revisit the behavior of atoms under
acceleration of the lattice, already studied
in~\cite{Morsch01,EarlyBlochoscillation,Denschlag02}, using the
adiabatic ramp-down analysis described in the last section. For
this particular experiment, we again decrease the mean oscillation
frequency of the magnetic trap to 19 Hz before turning the trap
off. Starting with the condensate at rest in the lab frame, we
linearly turn on the stationary lattice intensity over 40 \us\ in
order to ensure adiabaticity. The final depth for this experiment
is $ V = 13 \Er$. All the atoms are now approximately in the state
$|n= 0, q=0\rangle$. We then accelerate the lattice for 400 \us\
up to a given velocity $v_{\rm f}$, with a constant acceleration
$a \leq 800\ \rm{m}/\rm{s}^2$. The quasimomentum $q$ of the atoms
evolves during the acceleration according to a lattice version of
``Newton's law" $\dot{q} = -M a$~\cite{Ashcroft}. In the lattice
frame this is equivalent to adding a linear potential $-M a x$.
Provided that $|Ma\langle 1, q |x|0, q\rangle | \ll E_1-E_0$
\footnote{In the case of this 13\Er\ deep lattice, $E_0(q)$ and
$E_1(q)$ are almost independent of $q$.}, there is no transition
between the first two bands and the atoms stay in the lowest band.
This implies that the acceleration should be smaller than $4\times
10^4$ m/$\rm{s}^2$, a condition well satisfied in our experiment.
This acceleration allows us to produce any $q$ in the lowest band.
We note that combined with the loading in a moving lattice
described in section~\ref{dragging_moving_lattice} we can
therefore prepare the atoms in any Bloch state $|n,q\rangle$.

\begin{figure}
\includegraphics[width=8cm]{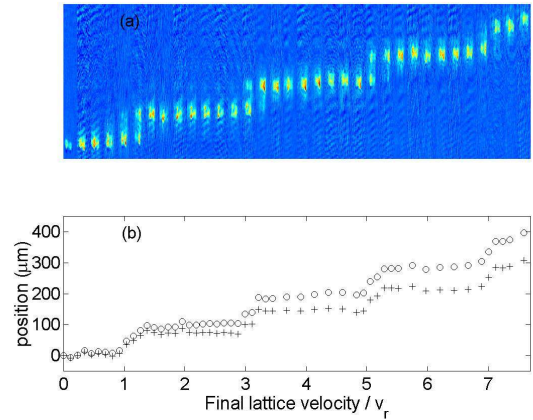}
\caption{(Color) Acceleration of atoms in the ground state ($n=0$)
band starting from $q = 0$. The lattice is adiabatically raised up
to $13 \Er$, accelerated and then adiabatically turned off at
constant velocity. Fig.~\ref{dragground20010905}a shows images of
the condensate after time of flight for increasing final lattice
velocities. Fig.~\ref{dragground20010905}b shows the position of
the center of mass of the cloud in the lab frame. The circles are
the positions measured on fig.~\ref{dragground20010905}a, whereas
the crosses represent the position of the cloud minus the
displacement due to the dragging of the lattice. It therefore
gives the momentum of the atoms.  } \label{dragground20010905}
\end{figure}

At the end of the acceleration period we ramp down the intensity
of the lattice in 200 \us, while still moving at $v_{\rm f}$.
After a 1.2 ms time-of-flight we take an absorption image of the
cloud. A series of pictures corresponding to different final
lattice velocities is shown in fig.~\ref{dragground20010905}a.

Those pictures show that if the final lattice velocity remains
within the first Brillouin zone (that is $|v_{\rm f}|< \vrec$) the
cloud comes back to rest in the laboratory frame after the
adiabatic ramping down of the lattice. This behavior is now well
understood in light of section~\ref{momentum_analysis}. On the
other hand, each time the lattice final velocity reaches $(2 m +
1)\vrec$ ($m$ being an integer) the atom momentum, after
ramping-down the lattice, in the lab frame, increases by steps
measured to be around $2\hbar k$. This momentum remains constant
for any lattice velocity between $(2 m + 1)\vrec$ and $(2 m +
3)\vrec$.

As another way to understand this behavior in the first Brillouin
zone, we again note that when the lattice, moving with constant
speed $v_{\rm f}= - q_{\rm f}/M$, is ramped down adiabatically the
velocity of the atoms with respect to the lattice varies from
$\frac{d E_0}{dq}(q_{\rm f})$ to $q_{\rm f}/M$ when the depth of
the lattice goes to 0. The velocity of the atoms in the lab frame
is thus $q_{\rm f}/M+v_{\rm f}= 0$.

On the other hand, if the final velocity is, say, between $\vrec$
and $3\vrec$, the velocity of the atoms in the lattice frame is no
longer $q_{\rm f}/M$ after ramping down the lattice, but $(q_{\rm
f} +2\hk)/M$. For example if the lattice is accelerated to $v_{\rm
f} = 2.5$\vrec, on ramping down, the velocity in the lattice frame
is $-0.5\vrec$. When the depth of the lattice approaches 0, the
velocity of the cloud in the lab frame thus goes to $(q_{\rm
f}+2\hk)/M-q_{\rm f}/M = +2\vrec$. This explains the jump in
momentum observed each time the velocity of the lattice reaches an
odd number of recoil velocities. As in
section~\ref{dragging_moving_lattice}, the final momentum is
independent of the intermediate lattice intensity. We have
repeated the experiment for $V = 1.5\Er$, $5\Er$ and $ 8\Er$ and
found exactly the same behavior, apart from the small
non-adiabaticity at the edge of a Brillouin zone. In the case of a
shallow lattice, we interpret this jump in momentum in the
laboratory frame as a first order Bragg diffraction: when the
velocity of the lattice reaches an odd  integer multiple of
$\vrec$ thus statisfying the Bragg condition, the momentum in the
lab frame changes by $2\hk$ in the same direction as the
acceleration. This Bragg diffraction is evidenced by the fact that
the state of the atoms in the lattice connects to a different free
parabola when the lattice is ramped down, as seen in
fig.~\ref{bandstructure4Er}.

One should not be misled by the fact that the condensate is back
at rest in the laboratory frame when $|v_{\rm f}|< \vrec$. The
cloud has been displaced, dragged along by the lattice. The
displacement is given by
\begin{equation}
x = \int_0^{\tau} \left(\frac{dE}{dq}(q(t))- q(t)/M\right) dt\ ,
\end{equation}
where $\tau$ is the duration of the experiment (600 \us). For
lattices deeper than about $3\Er$, the derivative almost vanishes
and we approximate the displacement by $x = v_{\rm f}\,
(\frac{1}{2}\tau_{\rm{accel}}+\tau_{\rm{ramp down}})$. In order to
determine whether the observed jump in momentum is exactly $2\hk$
at the crossing of the edge of the Brillouin zone, one has to
subtract this displacement due to the dragging of the lattice.
This is shown in fig.~\ref{dragground20010905}b. The circles are
the actual positions of the center of the cloud in the lab frame.
The crosses represent the positions corrected by the displacement
due to the dragging. The dispersion of the data on a given plateau
is due to a fluctuation of the position and velocity of the
condensate, with RMS values of about, respectively, 10 \um\ and
$0.03 \vrec$.

\begin{figure}
\includegraphics[width=8cm]{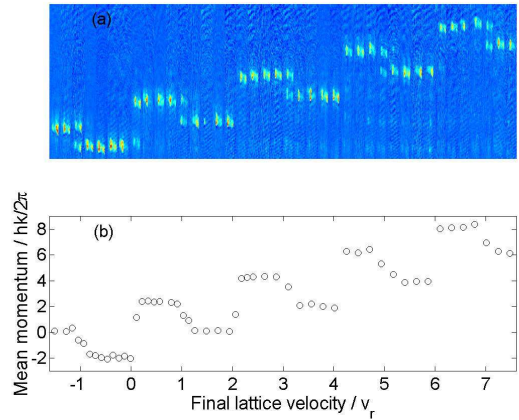}
\caption{(Color) Acceleration of the lattice with atoms initially
loaded into the band $n=1$, starting from $q = 1.5\hk$. The 13\Er\
lattice is raised adiabatically, accelerated, and then turned off
adiabatically. Fig.~\ref{dragfirst20010905}a shows the images of
the cloud in the lab frame after a 1.2 ms time of flight for
various final lattice velocities. Fig.~\ref{dragfirst20010905}b
presents the momentum deduced from the positions of the cloud,
after correction for the dragging. }\label{dragfirst20010905}
\end{figure}

We next consider essentially the same experiment except that we
now load the condensate in a lattice already moving with an
initial velocity $v_i = -1.5 \vrec$. Referring to
figure~\ref{bandstructure4Er} we see that adiabatic loading (100
$\mu$sec) prepares the atoms in the Bloch state
$|n=1,q=1.5\hk\rangle$. When the lattice is accelerated for 400
$\mu$sec in the positive direction in the lab frame, the atoms
follow the first band and the quasimomentum in the lattice frame
decreases linearly with time. Figure~\ref{dragfirst20010905}a
shows the position of the cloud in the lab frame after the
adiabatic ramp down of the lattice  (100 $\mu$sec) and the
subsequent 1.2 ms time of flight. In
figure~\ref{dragfirst20010905}b we show the average momentum of
the cloud. This includes compensation for the dragging of the
atoms during the time the lattice is on (600 $\mu$sec), as
described earlier.
Figure~\ref{dragfirst20010905}b shows an alternation of $-2\hk$
and $+4\hk$ momentum jumps in the lab frame. According to the
interpretation in terms of Bragg diffraction, when the final
velocity of the lattice reaches $ -\vrec$ (or the quasimomentum
reaches $+\hk$), the atoms undergo a first order Bragg diffraction
in the direction opposite to the acceleration of the lattice in
the lab (equivalently they change from the $0\hk$  to the $+2\hk$
free parabola, see figure 1). After being adiabatically released
from the lattice, they now travel at $-2\vrec$ in the lab frame,
in the direction opposite to the acceleration. Despite the lattice
being constantly accelerated in the direction of positive
momentum, at this stage the atoms gain a momentum in a direction
opposite to the acceleration. Further acceleration to $v_{\rm
f}=0$ leads to a second-order Bragg reflection that gives an
impulse of $+4\hk$, in the direction of the acceleration
(corresponding to a change from the $+2\hk$ parabola to the
$-2\hk$ parabola in the lattice frame, see fig.1). After adiabatic
release the atoms' momentum in the lab frame is $+2\hk$.

\begin{figure}
\includegraphics[width=8cm]{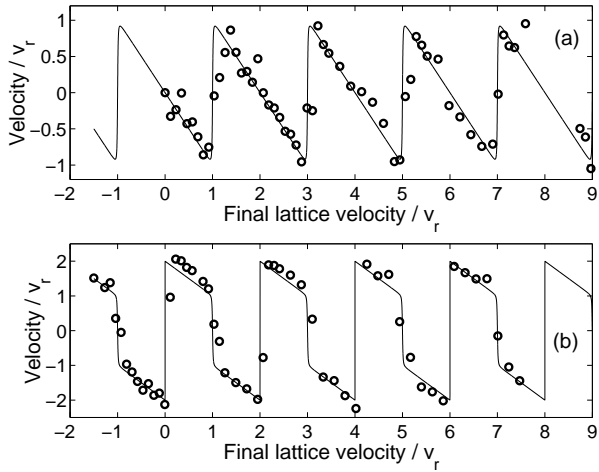}
\caption{Same experiment as in figure~\ref{dragground20010905} and
fig.~\ref{dragfirst20010905}. Velocity of the atoms in the lattice
frame after acceleration of the lattice, for different final
lattice velocities. The velocities are deduced from
fig.~\ref{dragground20010905}b and~\ref{dragfirst20010905}b by
subtracting the velocity of the lattice. In
fig.~\ref{velocity_blochoscill}a the atoms are prepared in the
ground state band, whereas they are prepared in the first band for
the results of fig.~\ref{velocity_blochoscill}b. The plain lines
are the theoretical group velocity calculated for a 0.1\Er\ deep
lattice.}\label{velocity_blochoscill}
\end{figure}

As a conclusion of this section, we discuss the difference between
the experiments presented here and earlier experiments
investigating Bloch oscillations. In reference~\cite{Morsch01},
for example, using the sudden turn-off method, the authors present
the variation of the mean velocity of the atoms in the lab frame,
after having accelerated the lattice. Their figure 2a shows steps
of amplitude $2\vrec$ (note that their $V_{\rm B} = 2 \vrec$). The
sharpness of the steps depends on the depth of the lattice and
becomes more gentle when the lattice gets deeper (see their figure
2c) and figure 13 (upper panel) of~\cite{Denschlag02}). On the
other hand our figure~\ref{dragground20010905}b exhibits sharp
steps very similar to figure 2a of \cite{Morsch01} taken with a
0.29\Er\ deep lattice, despite the fact we were using a 13\Er\
deep lattice, the same as fig. 13 of~\cite{Denschlag02}. The
adiabatic turn off method with any depth lattice thus produces
results equivalent to the sudden turn off method using a
vanishingly small lattice depth. This is because when we turn the
lattice intensity off adiabatically the states connect
continuously to the Bloch states for a vanishingly shallow
lattice. Figure~\ref{velocity_blochoscill}a shows the velocity of
the atoms with respect to the lattice when the lattice is off.
This is equivalent to figure 2b of \cite{Morsch01}, with even
sharper transitions.  Our transitions are nevertheless not
infinitely sharp because we are not adiabatic very close to the
zone boundary, as explained earlier.

Based on this discussion and the data of
figure~\ref{dragfirst20010905}  we can infer what Bloch
oscillations would look like in a weak lattice for a Bloch state
in the first excited  band. The velocity of the atoms in the
lattice frame is presented in figure~\ref{velocity_blochoscill}b.
This figure was again obtained from figure
~\ref{dragfirst20010905}b by subtracting the velocity of the
lattice. Note that in contrast to the usual Bloch oscillations in
the lowest band, here the Bloch oscillations in the first excited
band consist of a series of first and second order Bragg
diffractions, at integer multiples of \hk\ (half a reciprocal
lattice vector) each reversing the velocity in the lattice frame.
The first order Bragg diffraction changes the velocity in the
direction of the force acting on the atoms in the lattice frame,
whereas the second order Bragg diffraction changes the velocity by
twice as much in the direction opposite to the force. This is in
contrast with Bloch oscillations in the ground band where Bragg
reflection occurs at multiples of $2\hk$ (one reciprocal lattice
vector), always in the direction opposite to the force.

\section{Acceleration of atoms with a broad quasimomentum distribution}\label{incoherent atoms}

In a last set of experiments, we investigate the behavior of the
atoms under acceleration of the lattice when the atoms do not
occupy a single quasimomentum, but rather have a wide spread of
quasimomenta.

\begin{figure}
\includegraphics[width=8cm]{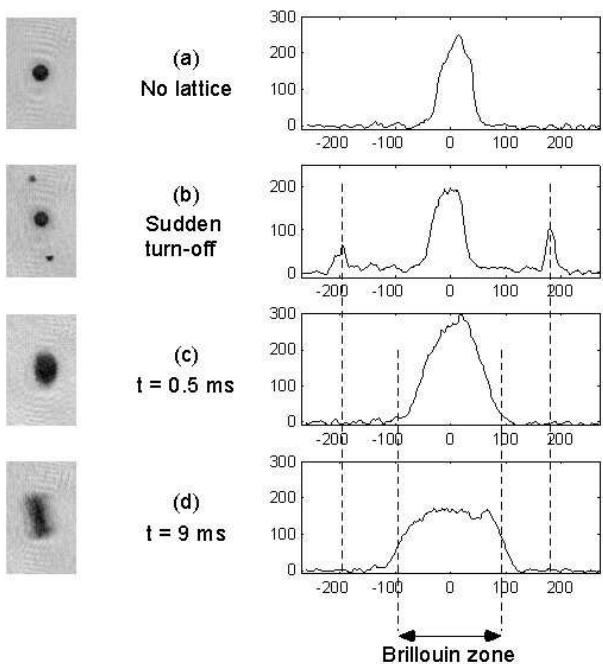}
\caption{Dephasing of a condensate sitting in a 5\Er\ deep lattice
for a time $t$. The time of flight has the same 3 ms duration for
all the images. The right column shows the density profiles of the
images in the left column, integrated perpendicular to the axis of
the lattice. In fig.\ref{decoherencesum}a, no lattice was applied.
In fig.\ref{decoherencesum}b the lattice was suddenly switched
off. In~\ref{decoherencesum}c and~\ref{decoherencesum}d the cloud
stays in the lattice for 0.5 and 9 ms respectively.}
\label{decoherencesum}
\end{figure}

\begin{figure}
\includegraphics[width=7cm]{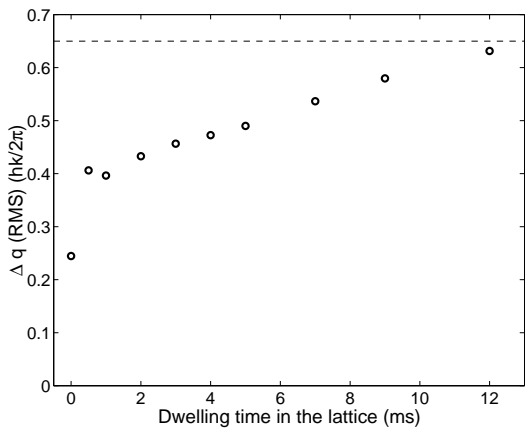}
\caption{ Evolution of the rms value of the observed the
quasimomentum distribution of the atoms  as a function of the time
spent by the atoms in the lattice. The dotted line represents the
expected width when we convolve the distribution of
fig.~\ref{decoherencesum}a with a uniformly populated first
Brillouin zone.} \label{Deltaqrms}
\end{figure}

In order to prepare a broad distribution of quasimomenta, we first
reproduce the experiment of~\cite{Greiner2001}: while the magnetic
trap is still on at a relatively high mean oscillation frequency
of 100 Hz in order to increase the interaction strength, we
adiabatically turn on a 5\Er\ deep lattice in 300 \us. We then
suddenly turn off the magnetic trap \footnote{The magnetic field
is turned off in about 100 \us. Furthermore the lattice beams are
red detuned in order to provide radial confinement in the absence
of the magnetic trap.} and let the atoms sit in the lattice for a
duration ranging from 100 $\mu$s to 12 ms. We follow the evolution
of the quasimomentum distribution of the atoms in the lattice by
adiabatically turning off the lattice (in 300 $\mu$s) and taking
an absorption image of the cloud after a 3 ms time of flight. The
results are shown in fig.~\ref{decoherencesum}:
\ref{decoherencesum}a shows an image of the undisturbed condensate
after the time of flight, as well as the density profile along the
lattice direction, integrated along the perpendicular direction,
with no lattice having been applied; in~\ref{decoherencesum}b, the
lattice has been switched off suddenly, immediately after
adiabatic loading of the lattice~\footnote{The width of the
central peak in fig.~\ref{decoherencesum}b is not due to the
quasimomentum spread before release. The mean field repulsion
between atoms increases the momentum spread after release. This
momentum spread is responsible for most of the observed width.}.
The momentum components at $\pm 2\hk$ appear and provide a
calibration of the scale; in the two last
pictures,~\ref{decoherencesum}c and ~\ref{decoherencesum}d, the
condensate sits for respectively $t = 0.5$ ms and $t = 9$ ms in
the lattice after which the lattice is ramped down in 300 \us.
After 9 ms of evolving time, when ramping down the lattice, the
momentum distribution of the cloud looks essentially like a
convolution of the  profile of figure~\ref{decoherencesum}a with
an almost uniformly populated momentum distribution with a $2\hk$
width. This corresponds to an almost completely filled first
Brillouin zone.
More quantitatively, from the integrated profiles we calculate the
rms width of the momentum distribution of the atoms in the lab
frame, after ramping down the lattice (see
figure~\ref{Deltaqrms}). Since the observed distributions result
from convolution of the quasimomentum distribution with the
distribution represented by fig.\ref{decoherencesum}a, we could in
principle deconvolve them in order to get only the contribution of
the quasimomentum. Instead, as a reference, we show in
fig.~\ref{Deltaqrms} the expected rms width, convolving the
experimental distribution of fig.~\ref{decoherencesum}a with a
quasimomentum distribution filling the first Brillouin zone.

An explanation for this broadening comes from the mean field
inhomogeneity across the cloud. In the magnetic trap, the chemical
potential is independent of the position. When the lattice is
super-imposed onto the magnetic trap, this is roughly still the
case, provided that the lattice is not too deep. When the magnetic
trap is suddenly turned off, the magnetic energy no longer
compensates for the mean field energy and the chemical potential
varies quadratically along the direction of the lattice. The rate
of change of the phase difference between two neighboring sites
then varies linearly along the lattice direction. This
inhomogeneity of the density across the condensate results in a
different phase evolution at each lattice site and consequently in
an effective dephasing of the single particle wave-function.
Remembering that the quasimomentum characterizes the phase
difference from one site to another, the apparent randomization of
this phase leads to a broadening of the quasimomentum
distribution. Roughly speaking, when the phase difference between
adjacent sites at the edge of the condensate reaches $2\pi$, the
wave function of an atom looks dephased, meaning that it is a
superposition of all quasimomenta in the first Brillouin zone.
This phase difference between neighboring sites at the edge of the
condensate being on the order of $\mu\, t/(N\hbar)$ (where $2N+1$
is the total number of sites), after a time evolution of duration
$t$, the time scale for dephasing is $\sim N\, h/\mu $.
Numerically, the condensate had a chemical potential $\mu/h = 5$
kHz. The estimated dephasing time (the time to create a $2\pi$
phase difference between adjacent wells) is then 8 ms, which is
approximately equal to the observed time to fill the Brillouin
zone. This treatment neglects tunnelling between lattice sites,
which would tend to equalize the phases. However, we calculate a
tunnelling rate of $2\pi \times 1.5$ kHz for a 5\Er\ deep lattice,
that is, the well-to-well tunnelling rate is faster than the
differential well-to-well phase evolution. Therefore, our simple
picture of dephasing is questionable, although it seems to give a
reasonable description of the experiment. We believe this point
deserves further study.(A more detailed study of some aspects of
mean-field dephasing in a lattice has been performed
in~\cite{Morsch03}.)

As an additional, albeit equivalent, demonstration  for the
randomization of the phase, we  look for diffraction  after
letting the condensate sit for a period of time. When we suddenly
turn off the lattice, we do not observe resolved diffraction peaks
when the atoms have spent more than 2 ms in the lattice. We
conclude that even though we may not have uniformly filled the
Brillouin zone after 2 ms, we broaden the quasimomentum
distribution sufficiently  that diffraction is not evident. We
note that in reference~\cite{Hadzibabic04}, the authors saw a
diffraction pattern from an array of about 30 independent
condensates. The difference is in their smaller number of lattice
sites, and may  also be influenced by differences in experimental
details such as the optical resolution for observing the
diffraction pattern, the number of diffraction peaks, and the fact
that the diffraction of ref.~\cite{Hadzibabic04} appears not to be
observed in the ``far field"~\footnote{By ``far field", we mean
that the diffraction pattern is observed after a time during which
the velocity of the diffracted momentum components moves them a
distance greater than the initial size of the cloud.}. We apply
the theory of~\cite{Hadzibabic04} to our about 80 interfering
condensates (assuming they are indeed independent which is only
partially valid in our case) and found no diffraction pattern, in
agreement with our observations.

\begin{figure}
\includegraphics[width=8cm]{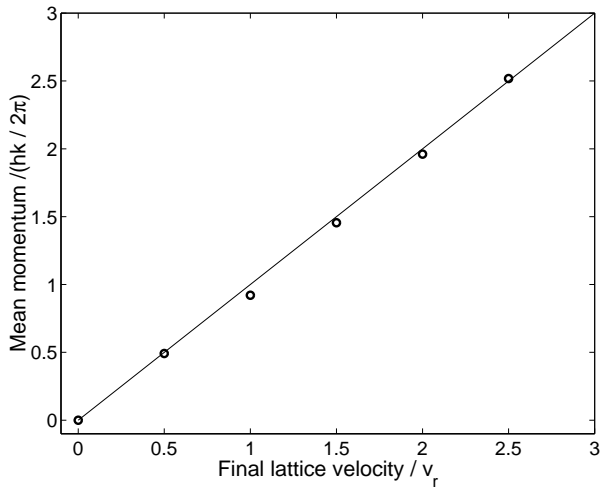}
\caption{Acceleration of a cloud of incoherent atoms in a 5\Er\
deep lattice. The circles represent the momentum of the center of
mass of the cloud, in the lab frame, after acceleration of the
lattice. The adiabatic ramp down is followed by a 3 ms time of
flight. The line represents the velocity of the lattice. }
\label{dragdephaseBEC}
\end{figure}

We finally turn to the behavior of the dephased cloud under
acceleration. After letting the cloud sit in a 5\Er\ deep lattice
for 5 ms, more than a sufficient time to broaden the quasimomentum
distribution enough that the diffraction is unresolved, we
accelerate the lattice in 500 $\mu$s to a chosen final velocity.
After this acceleration period, we ramp down the lattice depth to
zero in 100 $\mu$s and allow for a 3 ms time of flight, as
described earlier in the paper. The results are presented in
fig.~\ref{dragdephaseBEC}. In this figure, the mean momentum of
the cloud after ramping down the lattice intensity shows no sign
of the plateaus seen in fig.~\ref{dragground20010905}. This mean
momentum is proportional to the lattice velocity, in contrast to
the behavior described  in
section~\ref{single_bloch_acceleration}. In fact the atoms are
dragged at the velocity of the lattice, which means that in the
frame of the lattice their motion is frozen.

We reconcile this more intuitive behavior with the odd behavior of
section~\ref{single_bloch_acceleration} by assuming that the first
Brillouin zone is completely filled, and considering a small
component of the quasimomentum distribution, centered around
$q_0$. Upon acceleration, this population does undergo Bragg
diffraction when the velocity of the lattice reaches $q_0 + \hbar
k$, and exhibits the same step behavior as the one seen in
figure~\ref{dragground20010905}, the only difference being that
the horizontal axis is shifted by $q_0$. All the other
quasimomentum components are also Bragg reflected but at different
velocities of the lattice. When one averages the different
``staircase" patterns like the one shown in
fig.~\ref{dragground20010905} for all the quasimomenta in the
first Brillouin zone, the average velocity in the lab frame is the
velocity of the lattice. Another way to understand this is to
calculate the average group velocity for a uniformly populated
first Brillouin zone. That velocity is proportional to the average
of the slope of the $E_0(q)$. As the band is symmetric with
respect to $q = 0$, this average velocity with respect to the
lattice is zero and there is no motion of the center of mass of
the cloud with respect to the lattice.

We now compare our above results with two recent experiments
looking at thermal bosons and degenerate fermions in an optical
lattice~\cite{Cataliotti01,Modugno03}. In~\cite{Cataliotti01}, a
condensate surrounded by its thermal cloud is created in a several
\Er-deep lattice and a magnetic trap. The center of the magnetic
trap is then shifted and the subsequent behavior of the two
components is monitored. The authors observed that whereas the
thermal component is pinned and does not oscillate in the magnetic
trap, the condensate does oscillate with an oscillation frequency
modified
by the effective mass of the atoms in the lattice. 
The authors proposed an explanation based on the superfluidity of
the condensate that allows it to move through the corrugated
potential created by the lattice, whereas the thermal cloud does
not move due to its non-superfluid nature. In light of the
experiment we described above in this section, we propose an
alternate explanation for those results, based only on
single-particle band structure theory. In the experiment
of~\cite{Cataliotti01}, the condensate is prepared directly in the
lattice, and occupies the Bloch state $|n=0,q = 0\rangle$. Let us
now assume that the thermal component has a temperature that
corresponds to an energy between the ground state band and the
first band of the lattice. The ground  band is then almost
uniformly populated, meaning that the single-particle
wave-function of the atoms effectively contains all quasimomenta
in the first Brillouin zone. Shifting the position of the magnetic
trap is equivalent to accelerating both the lattice and the trap
with respect to the lab and therefore equivalent to applying a
uniform force to the atoms. As described above, the thermal cloud
filling the Brillouin zone does not move  with respect to the
lattice (figure~\ref{dragdephaseBEC}). This is what the authors of
reference~\cite{Cataliotti01} observed and it is completely
consistent with a single-particle description, without any
reference to superfluidity or critical velocity, phenomena
dependent on interactions.

We finally discuss briefly the recent experiment dealing with
degenerate fermions in a one-dimensional optical
lattice~\cite{Modugno03,Roati04}. In reference~\cite{Modugno03}, a
Fermi sea of $^{40}\rm{K}$ is produced in an optical lattice. The
authors observe the absence of peaks in the diffraction pattern
after sudden turn off of the lattice. This implies that the Fermi
momentum  is comparable to or larger than $\hbar k$ so that the
quasimomentum extends throughout the Brillouin zone, similar to
our dephased cloud of Bosons. In the same work the authors repeat,
with the Fermi gas, the experiment of
reference~\cite{Cataliotti01} where they shift the magnetic trap
with respect to the lattice. Consistent with our single-particle
interpretation of the experiment (and with the single particle
interpretation given in~\cite{Modugno03}), they do not observe
oscillation of the Fermi cloud in the magnetic trap.
In~\cite{Roati04}, the authors again produce the Fermi sea in a
lattice but this time they only partially fill the first Brillouin
zone. As a result they do observe a diffraction pattern consisting
of resolved peaks when suddenly releasing the atoms from the
lattice. They also observe Bloch oscillations in their vertical
lattice, due to gravity, as should be the case for a partially
filled Brillouin zone. As ultra-cold indistinguishable fermions
are essentially non-interacting (no $s$-wave collisions), the
experiments of~\cite{Modugno03,Roati04} illustrate single
particle, i.e. non-interacting particle, behavior of a cloud of
cold atoms in a lattice, as those authors point out. Collisions
imply a coupling between quasimomentum states and thus the failure
of the single particle (Bloch theory) description. The behavior of
ultra-cold fermions is identical to our experiment and the
experiment of \cite{Cataliotti01} with interacting bosons, when
the influence of interactions is negligible on the time scale of
the experiment. It is particularly striking that Fermions and
Bosons can behave exactly in the same way under some
circumstances: all that matters is the way the Brillouin zone is
filled, although the way this filling occurs may depend on the
quantum statistics.

\section{Conclusion}

In summary, we have presented a series of experiments in which a
condensate is adiabatically loaded into an optical lattice,
preparing the atoms in a single Bloch state. In a first set of
experiments, the lattice is initially moving, and the atoms come
back to rest in the lab frame after the adiabatic turn off of the
lattice, leading to non-intuitive behavior for this ``quantum
conveyor belt". In a second set of experiments, we act on the
prepared quasimomentum distribution by accelerating the lattice.
We then analyze the new quasimomentum distribution by again
adiabatically ramping down the lattice, and again observed
non-intuitive behavior. We observe discrete jumps in the resulting
momentum distribution, depending upon the velocity of the lattice.
These jumps are reminiscent of Bragg diffraction at each avoiding
crossing due to the laser coupling and are equivalent to Bloch
oscillations. In a last set of experiments, we let the initial
quasimomentum distribution evolve under the influence of
interactions between the atoms, leading to the filling of the
first Brillouin zone. The resulting cloud now exhibits a different
behavior under acceleration of the lattice, i.e. the cloud appears
to be frozen in the frame of the lattice. Finally we showed the
similarities between the behavior of a cold thermal cloud and that
of a cloud of degenerate fermions in an accelerated optical
lattice, when the quasimomentum extends throughout the first
Brillouin zone.

Among the issues that we believe deserve further studies, both
experimentally and theoretically, is the competition between phase
winding and tunnelling, that is to say how atoms lose their
well-to-well phase coherence. Furthermore, we considered in this
paper the dephasing of the wave-function due to the density
profile of the cloud, but the quantum fluctuation of the atom
number in each well is also a source of effective decoherence that
should be explored. We also emphasize that the time scales of our
experiments are very short with respect to other experiments, such
as the one described in~\cite{Morsch03,Cataliotti01}.

Finally we note that the method of
section~\ref{single_bloch_acceleration} could be useful for
precision measurements, as described in ref.~\cite{Battesti04}.

\begin{acknowledgments}
We are pleased to thank P. B. Blakie for enlightening discussions.
We acknowledge funding support from the US Office of Naval
Research, NASA, and ARDA/NSA. H.H. acknowledges funding from the
Alexander von Humboldt foundation.
\end{acknowledgments}


\begin{thebibliography}{30}
\bibitem{Anderson} M.H. Anderson, J.R. Ensher, M.R. Matthews, C.E.
Wieman, E.A. Cornell, Science {\bf 269}, 198 (1995).

\bibitem{dafolvo99} F. Dafolvo, S. Giorgini, L. P. Pitaevskii, and S. Stringari, Rev. Mod. Phys. {\bf
71}, 463 (1999).

\bibitem{Denschlag02} J. Hecker-Denschlag, J. E. Simsarian, H. H\"affner, C. McKenzie, A. Browaeys, D. Cho,
K. Helmerson, S. L. Rolston, and W. D. Phillips,  J. Phys. B {\bf
6}, 1 (2002).

\bibitem{Fallani03} L. Fallani, F.S. Cataliotti, J. Catani, C.
Fort, M. Modugno, M. Zawada, and M. Inguscio, Phys. Rev. Lett.
{\bf 91}, 240405 (2003).

\bibitem{Morsch01} O. Morsch, J.H. M\"{u}ller, M. Cristiani, D.
Ciampini, and E. Arimondo, Phys. Rev. Lett. {\bf 87} 140402
(2001).

\bibitem{Eiermann03} B. Eiermann, P. Treutlein, T. Anker, M. Albiez, M. Taglieber, K.-P. Marzlin,
and M. K. Oberthaler, Phys. Rev. Lett. {\bf 91}, 060402 (2003).

\bibitem{Kozuma99} M. Kozuma, L. Deng, E.W. Hagley, J. Wen, R.
Lutwak, K. Helmerson, S.L. Rolston, W.D. Phillips, Phys. Rev.
Lett. {\bf 82}, 871 (1999).

\bibitem{Ashcroft} N. Ashcroft and D. Mermin, ``Solid State Physics",
Saunders College (1976).

\bibitem{Petrich95} W. Petrich, M. H. Anderson, J. R. Ensher, and
E. Cornell, Phys. Rev. Lett. {\bf 74}, 3352 (1995).

\bibitem{Stenger99} J. Stenger, S. Inouye, A.P. Chikkatur, D.M. Stamper-Kurn, D.E. Pritchard, and W. Ketterle,
Phys. Rev. Lett. {\bf 82}, 4569 (1999).

\bibitem{Kastberg95} A. Kastberg, W. D. Phillips, S. L. Rolston, and R. J. C. Spreeuw,
Phys. Rev. Lett. {\bf 74}, 1542 (1995).

\bibitem{Greiner2001} M. Greiner, I. Bloch, O. Mandel, T. W. H\"ansch, T. Esslinger,
Phys. Rev. Lett. {\bf 87}, 160401 (2001).

\bibitem{EarlyBlochoscillation} M. Ben Dahan {\it et al.}, Phys. Rev. Lett.
{\bf 76}, 4508 (1996), S. R. Wilkinson {\it et al.}, Phys. Rev.
Lett. {\bf 76}, 4512 (1996).

\bibitem{Morsch03}  O. Morsch, J. H. M\"uller, M. Cristiani, P. B. Blakie, C. J. Williams, P. S. Julienne, and E. Arimondo,
Phys. Rev. A {\bf 67}, 031603 (2003).

\bibitem{Hadzibabic04} Z. Hadzibabic, S. Stock, B. Battelier, V. Bretin, and J. Dalibard, Phys. Rev. Lett. {\bf
93}, 180403 (2004).

\bibitem{Cataliotti01} F.S. Cataliotti, S. Burger, C. Fort, P.
Maddaloni, F. Minardi, A. Trombettoni, A. Smerzi, M. Inguscio,
Science {\bf 293}, 843 (2001); S. Burger, F. S. Cataliotti, C.
Fort, F. Minardi, and M. Inguscio, M. L. Chiofalo and M. P. Tosi,
Phys. Rev. Lett. {\bf 86}, 4447 (2001).

\bibitem{Modugno03} G. Modugno, F. Ferlaino, R. Heidemann, G. Roati, and M. Inguscio, Phys. Rev. A {\bf 68},
011601 (2003).

\bibitem{Roati04} G. Roati, E. de Mirandes, F. Ferlaino, H. Ott,
G. Modugno, and M. Inguscio, Phys. Rev. Lett. {\bf 92}, 230402
(2004).

\bibitem{Battesti04} R. Battesti, P. Cladé, S. Guellati-Khélifa, C. Schwob, B. Grémaud, F. Nez,
L. Julien, and F. Biraben, Phys. Rev. Lett. {\bf 92}, 230402
(2004).

\end{thebibliography}
\end{document}